\documentclass[amsmath,amssymb,prl,hyperlink,twocolumn]{revtex4}

\usepackage{graphicx}
\usepackage{soul}
\usepackage[colorlinks=true,citecolor=blue,linkcolor=magenta]{hyperref}
\usepackage[usenames]{color}
\usepackage{amsfonts}

\begin{document}

\title{On-demand semiconductor single-photon source with near-unity indistinguishability}

\author{Yu-Ming He$^1$, Yu He$^1$, Yu-Jia Wei$^1$, Dian Wu$^1$, Mete Atat\"{u}re$^{2,1}$, Christian Schneider$^3$, \\ Sven H\"{o}fling$^3$, Martin Kamp$^3$,  Chao-Yang Lu$^{1,2}$, Jian-Wei Pan$^1$\vspace{0.2cm}}

\affiliation{$^1$ Hefei National Laboratory for Physical Sciences at Microscale and Department of Modern Physics, University of Science and Technology of China, Hefei, Anhui 230026, China}
\affiliation{$^2$ Cavendish Laboratory, JJ Thomson avenue, University of Cambridge, CB3 0HE, Cambridge, United Kingdom}
\affiliation{$^3$ Technische Physik, Physikalisches Instit\"{a}t and Wilhelm Conrad R\"{o}ntgen-Center for Complex Material Systems, Universitat W\"{u}rzburg, Am Hubland, D-97074 W\"{u}zburg, Germany}

\date{\vspace{0.1cm}\today}

\begin{abstract}
Single photon sources based on semiconductor quantum dots offer distinct advantages for quantum information,
including a scalable solid-state platform, ultrabrightness, and interconnectivity with matter qubits.
A key prerequisite for their use in optical quantum computing and solid-state networks is a high level of efficiency and indistinguishability.
Pulsed resonance fluorescence (RF) has been anticipated as the optimum condition for the deterministic generation of high-quality photons with vanishing effects of dephasing.
Here, we generate pulsed RF single photons on demand from a single, microcavity-embedded quantum dot under \textit{s}-shell excitation with 3-ps laser pulses.
The $\pi$-pulse excited RF photons have less than 0.3$\%$ background contributions and a vanishing two-photon emission probability.
Non-postselective Hong-Ou-Mandel interference between two successively emitted photons is observed with a visibility of 0.97(2),
comparable to trapped atoms and ions. Two single photons are further used to implement a high-fidelity quantum controlled-NOT gate.
\end{abstract}

\pacs{78.67.Hc, 42.50.Dv, 42.50.St, 78.55. 42.50.Ar}

\maketitle

Single photons have been proposed as promising quantum bits (qubits) for quantum communication \cite{Pan-RMP}, linear optical quantum computing \cite{Kok-RMP,Obrien-Science} and as messengers in quantum networks \cite{Kimble-Nature2008}. These proposals primarily rely upon a high degree of indistinguishability between individual photons to obtain the Hong-Ou-Mandel (HOM) type interference \cite{HOM} which is at the heart of photonic controlled logic gates and photon-interference-mediated quantum networking \cite{Pan-RMP,Kok-RMP,Obrien-Science,Kimble-Nature2008}.

Among different types of single-photon emitters \cite{SPS2,SPS3}, quantum dots (QDs) are attractive solid-state devices since they can be embedded in high-quality nanostructure cavities and waveguides to generate ultra-bright sources of single and entangled photons \cite{SPS3,Atac-Science2000,Santori-PRL2001,Nature-Ultrabright2010France}. QDs also provide a light-matter interface \cite{Fushman-2008Science,Atac-PRL-interface,Young-2011PRA} and can in principle be scaled to large quantum networks \cite{LiuRenBao-PRL-network}. Two-photon HOM interference experiments using photons from a single QD \cite{Santori-Nature2002,Bennett-OptLett,Michler-PSS}, as well as from independent sources \cite{Glen-PRL,Patel-Nature Photonics}, have not only demonstrated the potential of QDs as single-photon sources, but also revealed the level of dephasing arising from incoherent excitation. The method of incoherent pumping (via above band-gap or \textit{p}-shell excitation) typically causes reduced photon coherence times due to homogeneous broadening of the excited state \cite{Bennett-OptLett} and uncontrolled emission time jitter from the nonradiative high-level to \textit{s}-shell relaxation \cite{footnote1}, leading to a decrease of photon indistinguishability.

To eliminate these dephasings, an increasing effort has been devoted to \textit{s}-shell resonant optical excitation of QDs. The Mollow triplet spectra and photon correlations of the resonance fluorescence (RF) have been measured \cite{Muller-PRL,Nick-NatPhys,Flagg-NatPhys,Michler-NatPhotonics}. Under continuous-wave (CW) laser excitation, a high degree of indistinguishability for continuously generated RF photons has been demonstrated through post-selective HOM interference \cite{Ates-PRL2009}. However, in the CW regime, as the emission time of the RF photons is uncontrolled, the HOM interference relies on the finite single-photon detection time resolution to discriminate and post-select a small fraction of photons that overlapped on the beam-splitter at the same time \cite{Kiraz-PRL,Patel-PRL,Ates-PRL2009}. Therefore, the obtained interference visibility needs to be convoluted with---and is thus limited by---the realistic detection time response. This limitation, together with the low efficiency of two-photon interference owing to the unsynchronized photon arrival time, prohibits the direct application of CW RF photons in many quantum information protocols \cite{Pan-RMP,Kok-RMP,Obrien-Science,Kimble-Nature2008}. More recent experiments operating on the low excitation regime have showed that the coherent scattering part of the RF could have coherence comparable to the excitation laser \cite{Clemens-PRL,French-APL}. However, such a single-photon source would suffer an intrinsically low efficiency.

It has been anticipated \cite{Michler-PSS,Ates-PRL2009,footnote1,Patel-Nature Photonics,Glen-PRL} that \emph{pulsed} and resonant \textit{s}-shell excitation could remedy the above problems and be used for deterministic generation of time-tagged, highly indistinguishable single photons. In addition, the pulsed and transition-selective RF single photons are also a prerequisite for the much sought-after goal of entangling distant QD spins through photon interference \cite{Kimble-Nature2008, Barrett}, as well as for the scheme of generating on-demand multi-photon cluster states \cite{Terry09}. Earlier experiments \cite{Muller-PRL,Rabi4} have used pulsed resonant excitation to demonstrate Rabi oscillation, a hallmark for quantum optics. Yet, access to a background-free on-demand single-photon source with near-unity indistinguishability proved elusive \cite{add}.


In this Article, by applying resonant \textit{s}-shell optical excitation with picosecond laser pulses, we generate pulsed RF single photons on demand from a single QD embedded in a planar microcavity. Rabi oscillations are visible from the variation of the RF intensity as a function of pump pulse area. Under deterministic $\pi$-pulse excitations, the RF photons have less than 0.3$\%$ background contributions and show an anti-bunching of $g^2(0)=0.012(2)$. We observe non-postselective HOM interference with a raw visibility of 0.91(2) and corrected visibility of 0.97(2) for two RF photons excited by two successive $\pi$ pulses separated by 2ns. Finally, the highly indistinguishable RF photons are utilized to demonstrate a quantum controlled-NOT gate.

\subsection{Pulsed resonance fluorescence}
Our experiments are performed on self-assembled InGaAs QDs which are embedded in a planar microcavity and cooled in a cryogen-free bath cryostat at 4.2K (see Fig.$\,$S1). Laser excitation of a single QD and collection of the emitted fluorescence are carried out with a confocal microscope. The excitation laser is pulsed with nominal pulse width of 3ps. The microscope is operated in a cross-polarization configuration, whereby a polarizer is placed in the collection arm with its polarization perpendicular to the excitation light, extinguishing the scattered laser by a factor exceeding $10^6$. The microcavity has a quality factor of $\sim\,$200 which increases the fluorescence collection efficiency and reduces the laser power required for excitation of the QDs.

Figure$\,$1a shows the detected RF photon counts as a function of the square root of the excitation laser power. The oscillation of the RF intensity is due to the well-known Rabi rotation between the ground and the excitonic state. It has been demonstrated previously by quasi-resonant \cite{Rabi1,Rabi2,Rabi3} or resonant driving \cite{Muller-PRL,Rabi4}. The RF intensity reaches its first peak at the $\pi$ pulse. We excite the QD with $\pi$ pulses at a repetition rate of $\thicksim\,$82MHz and observed $\thicksim\,$230,000 photon counts on a single-photon detector (with an efficiency of 22$\%$). The overall RF collection efficiency is $\thicksim\,$1.3$\%$. After correcting for the fibre coupling efficiency ($\thicksim\,$45$\%$), polarizer ($\thicksim\,$50$\%$) and beam splitter ($\thicksim\,$95$\%$), we estimate that $\thicksim\,$6$\%$ of the photons emitted by the QD are collected into the first lens, which is in good agreement with numerical simulations (see Supplementary Information). To verify that it is indeed a single-photon source, Figure$\,$1b shows the second-order correlation measurement of the $\pi$-pulse driven RF photons. At zero delay, it shows a clear anti-bunching with a vanishing multi-photon probability of $g^2(0)=0.012(2)$. Thus it can be concluded that one and only one RF photon is generated on demand from every $\pi$-pulse excitation. However, the photon extraction efficiency needs to be drastically improved for it to become a deterministic single-photon source.

Figure$\,$2a shows a linear-log plot of the pulsed RF (the sharp central line) together with the residual laser leakage (the broadband feature fitted by the red line) monitored on a spectrometer. Taking advantage of the huge linewidth mismatch between the RF signal and the laser background, we pass the RF through an etalon which has a bandwidth of $\thicksim\,$20GHz---much wider than that of the RF photons and much narrower than that of the pulsed laser---to further suppress the excitation laser background. This results in a clean RF spectrum as shown in the inset of Fig.$\,$2a, with an improvement of the signal to background (including the detector dark counts) ratio from 20 to 357 at $\pi$-pulse excitation. For a range of laser powers, the signal to background ratio is extracted and plotted in Fig.$\,$2b.

\begin{figure}[tb]
    \centering
        \includegraphics[width=0.49\textwidth]{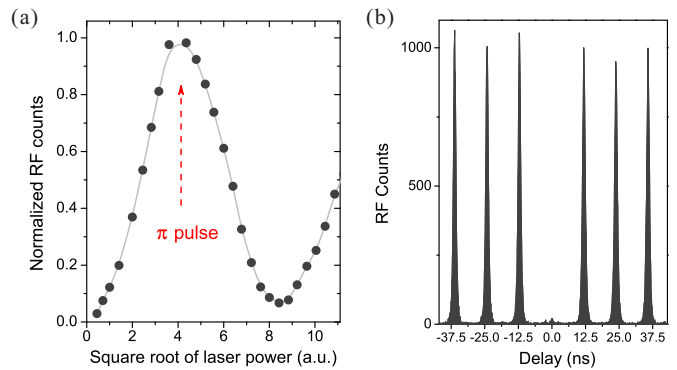}
\caption{Rabi oscillation and anti-bunching. (a) RF intensity versus square-root of incident power. The gray line is a guide to eye. While excitation-induced damping of the Rabi oscillation (there has been an intense debate on its mechanism, see e.g. \cite{ROdamping}) is visible at higher powers, a $\pi$-pulse excitation is obtained with reasonable quality. Our current work only focuses on the $\pi$-pulse regime. (b) Intensity-correlation histogram of the RF emission from the QD under pulsed \textit{s}-shell excitation obtained using a Hanbury Brown and Twiss-type setup. The second-order correlation $g^2(0)=0.012(2)$ is calculated from the integrated photons counts in the zero time delay peak divided by the average of the adjacent six peaks, and its error (0.002) which denotes one standard deviation, is deduced from propagated Poissonian counting statistics of the raw detection events.}
\label{fig:new1}
\end{figure}

\begin{figure*}[tb]
    \centering
        \includegraphics[width=0.975\textwidth]{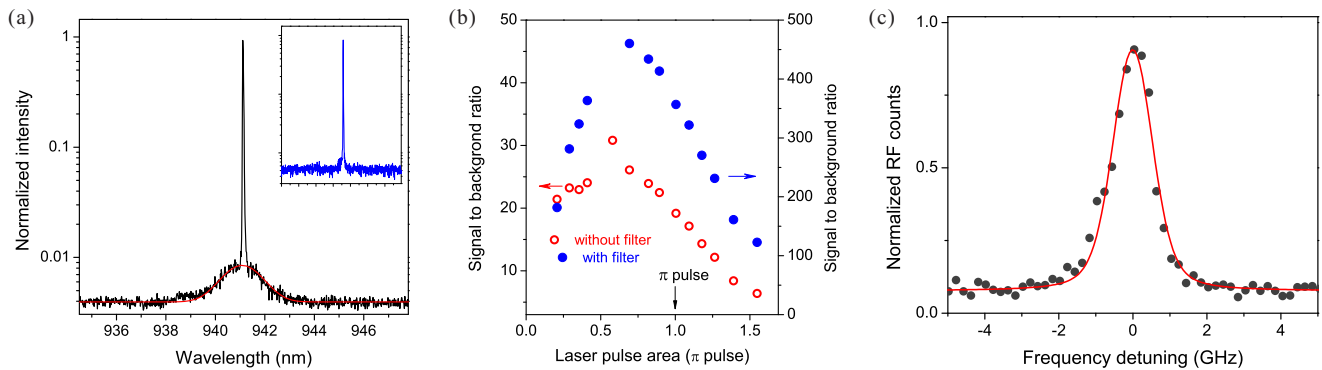}
\caption{Spectra of the pulsed RF. (a) An example of the pulsed RF displayed on a spectrometer and plotted using a linear-log scale. The excitation laser profile (fitted by the red line) is much broader than the RF signal, enabling a second-stage of background filtering based on frequency (see text) which is not possible in the CW case. The inset shows the spectrum after passing the RF photons through a 20-GHz etalon to eliminate the residual laser background. (b) Pulsed RF signal to background ratio for a range of excitation powers with and without filtering. (c) A high-resolution RF spectrum when excited by a $\pi$ pulse. The red line was fitted using a Voigt profile.}
\label{fig:3}
\end{figure*}

A typical example of high-resolution spectra of the pulsed RF measured using a Fabry-P\'{e}rot scanning cavity is shown in Fig.$\,$2c. It shows a pronounced deviation from the Lorentzian lineshape obtained from CW excitation as shown in Fig.$\,$S2, and can be fitted with a Voigt profile with a homogeneous linewidth of 0.4(1)GHz (corresponding to T$_2$=0.7(2)ns) and an inhomogeneous linewidth of 1.0(1)GHz. The spontaneous emission lifetime for this QD is measured to be T$_1$=0.41(2)ps (see Fig.$\,$S3), and we estimate the pure dephasing time T$^*_2$=5.7$^\infty_{3.8}\,$ns. The Gaussian component in this Voigt profile could potentially be caused by spectral diffusion owing to pulsed-laser-induced charge fluctuations in the vicinity of the QD (trapping and untrapping of charges in nearby defects and impurities) \cite{diffusion1,diffusion2,diffusion3}. The inhomogeneous linewidth varies for different QDs and typically shows an increase at larger excitation power (see Fig.$\,$S3), which is in qualitative agreement with previous investigations of light-induced spectral diffusion  \cite{diffusion1}.

\subsection{Two-photon quantum interference}
To perform pulsed two-photon interference, we adopt a similar experimental configuration (see Fig.$\,$3a) as in ref.$\,$\cite{Santori-Nature2002}. Each excitation laser pulse, originally separated by $\thicksim$12.5ns, is further split into two pulses with a 2-ns delay. Thus, every $\thicksim$12.5ns, the QD is excited twice, generating two successive single RF photons. The output RF photons are then fed into an unbalanced Mach-Zehnder interferometer with a 2-ns path-length difference (Fig.$\,$3a). The two outputs of this interferometer are detected by single-mode fiber-coupled single-photon counters, and a record of coincidence events is kept to build up a time-delayed histogram (for more details see Fig.$\,$S4).

Figure$\,$3(b) and (c) show the central cluster of the histogram when the two $\pi$-pulse excited single photons, before recombining in the last beam splitter, are prepared in cross and parallel polarization states respectively. The five peaks, from left to right, corresponds to the cases where the two photon arrives at the beam splitter with a time delay of -4ns, -2ns, 0ns, 2ns, and 4ns, respectively. For distinguishable photons with different polarization, the expected peak-area ratio equals 1:2:2:2:1, which is in good agreement with Fig.$\,$3b.
\begin{figure*}[tb]
    \centering
        \includegraphics[width=0.99\textwidth]{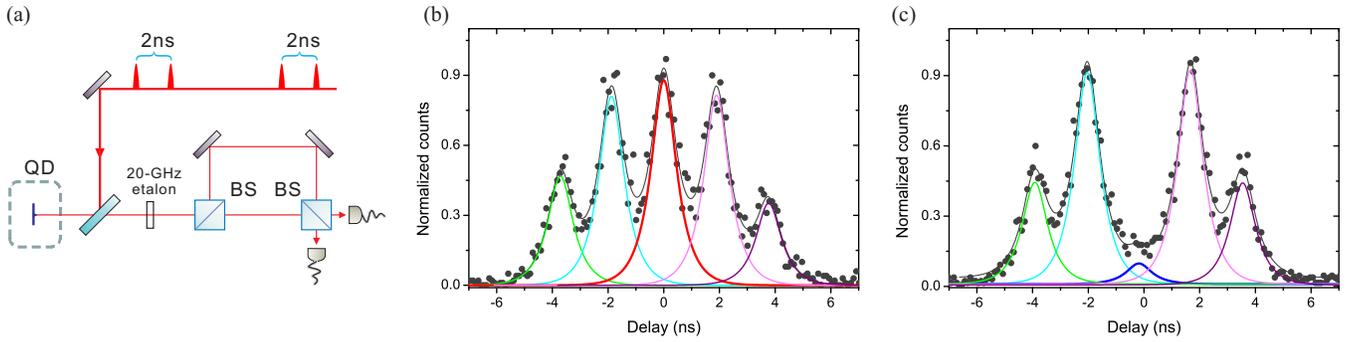}
\caption{Non-postselective HOM-type interference between two pulsed RF single photons. (a). Two unbalanced Mach-Zehnder interferometers with a path length difference of 2ns are used both in the excitation arm (not shown) and in the two-photon interference. (b-c). The central cluster of the histogram (see Fig.$\,$S4 for the full histogram) of two-photon detection events with a relative delay time. In (b) and (c), the input two photons are $\pi$-pulse excited and prepared in cross and parallel polarizations, respectively. The fitting function for each peak is the convolution of a double exponential decay (exciton decay response) with  a Gaussian (single-photon detector time response). Due to the limited time response, the five peaks have finite overlaps. The area of the fitted central peaks, covered by the red line in (b) and the blue line in (c), respectively, are extracted and used to calculate the visibility.}
\label{fig:4}
\end{figure*}

If two perfectly indistinguishable photons are superposed on a beam splitter, they will always exit the beam splitter together through the same output port, leading to a zero coincidence rate---the HOM dip \cite{HOM} which cannot explained by classical optics. Figure$\,$3c shows a strong suppression of the coincidence counts at zero delay when the two incoming photons are prepared in the same polarization state. Quantitative evaluation (see the caption of Fig.$\,$3 for details) shows that the probability of the two photons to exit the same channel in a 2-photon Fock state (bunching) is 95.4$\%$. This corresponds to a raw two-photon HOM interference visibility of 0.91(2).

Taking into account the residual two-photon emission probability $g^2(0)=0.012(2)$, and the optical imperfections of our interferometric setup which are independently measured, $R/T=1.01$ and $(1-\varepsilon)=0.98$, where $R$, $T$ are the reflectivity and transmitivity of the beam splitter and $(1-\varepsilon)$ is the first-order interference visibility of the Mach-Zehnder interferometer tested with a CW laser, we obtain corrected degrees of indistinguishability to be 0.97(2). The visibility can be further increased slightly by decreasing the excitation laser power. On another QD, we test the HOM interference with $\pi$, 0.72$\pi$ and 0.41$\pi$ pulse excitation and observe visibilities of 0.96(6), 0.97(6) and 0.99(4), respectively (see the data in Fig.$\,$S4).

Taken together, these are to date the highest visibilities reported for QD-based single-photon sources. These results demonstrate that the solid-state pulsed RF single photons in quick succession are highly indistinguishable to a level comparable to the best results from those well-developed systems such as parametric down-conversion \cite{Pan-RMP}, trapped atoms and ions \cite{atom1, atom4, atom2,atom3}. The high-visibility results indicate a reduction of the fast dephasing and an elimination of the emission time jitter associated with the pulsed RF, compared to the previous incoherent excitation methods. The pure dephasing time T$^*_2$=5.7$^\infty_{3.8}$ns is considerably larger than the 2ns and thus should have little effect on the visibility. The spectral diffusion (as shown in Fig.$\,$2c) should also happen at a time scale much longer than the 2-ns separation, which is consistent with previous experiments \cite{Santori-Nature2002,Bennett-OptLett,diffusion1,diffusion3}.

\subsection{Controlled-NOT gate with single photons}
We now demonstrate how the on-demand RF single photons can be utilized to implement a quantum controlled-NOT (CNOT) gate. The quantum CNOT gate is a fundamental two-qubit logic gate. If the control qubit is in logic $|0\rangle_c$, nothing happens to the target qubit, whereas if the control qubit is in logic $|1\rangle_c$, the target qubit will flip ($|0\rangle_t\rightarrow|1\rangle_t$, $|1\rangle_t\rightarrow|0\rangle_t$). The photonic CNOT gate is a basic building block for quantum computing and has been demonstrated many times with down-converted photons \cite{photonCNOT,photonCNOT2,photonCNOT3,photonCNOT4}, and very recently, with \textit{p}-shell excited single photons from QDs \cite{CNOT}.
\begin{figure*}[tb]
    \centering
        \includegraphics[width=0.98\textwidth]{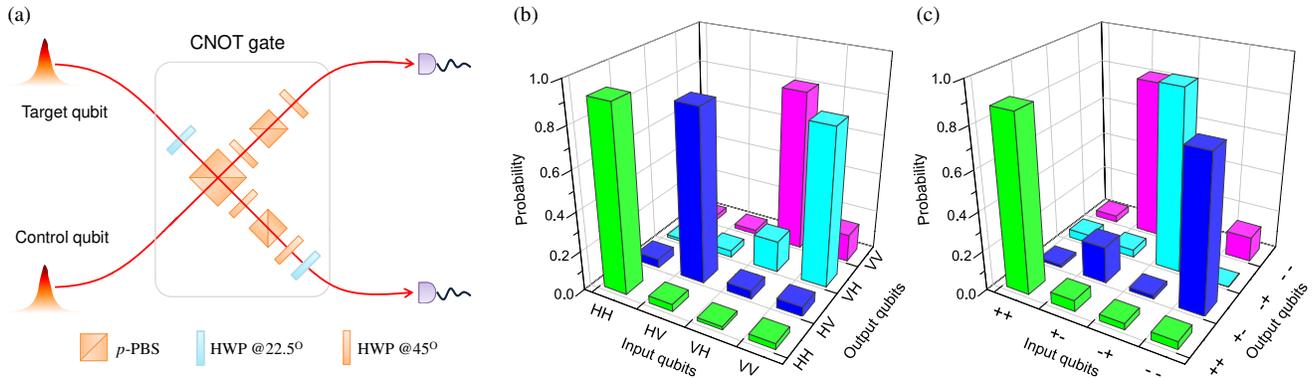}
\caption{Realization of a quantum CNOT gate using pulsed RF single photons. (a). The optical circuit. The control and target qubits are from the two successively emitted RF photons with a 2-ns delay. The half-wave plates (HWPs) placed at $\theta=45^\circ$ and $\theta=22.5^\circ$ are used to realize the unitary rotation: $\left|H\right\rangle\rightarrow \cos({2}\theta)\left|H\right\rangle+\sin({2}\theta)\left|V\right\rangle$, $\left|V\right\rangle\rightarrow \sin({2}\theta)\left|H\right\rangle-\cos({2}\theta)\left|V\right\rangle$. (b-c). The experimentally measured truth table. The coincident count rates are converted to probabilities by normalizing them with the sum of coincidence counts obtained for the respective input state. Ideally, the CNOT truth table in the $ZZ$ basis should gives a unity possibility for the input-output combination $HH\rightarrow HH$, $HV\rightarrow HV$, $VH\rightarrow VV$ and $VV\rightarrow VH$, and zero possibility for others. Similarly, in the $XX$ basis, there should be only  $++\rightarrow++$, $+-\rightarrow--$, $-+\rightarrow-+$ and $--\rightarrow+-$ ($\left|\pm\right\rangle=(1/\sqrt{2})(\left|H\right\rangle\pm\left|V\right\rangle)$). The unwanted combinations are mainly caused by the imperfections of the optical elements and the remaining distinguishability of the single photons.}
\label{fig:5}
\end{figure*}

We prepare two input qubits encoded in the polarization states of the pulsed RF single photons $\left| \alpha \right\rangle_t = a\left|H\right\rangle_t  + b\left|V\right\rangle_t$ and $\left| \beta\right\rangle_c = c\left|H\right\rangle_c+d\left|V\right\rangle_c $, where $H$($V$) denotes horizontal$\,$(vertical) polarization and is used to encode $0(1)$. The two inputs are then fed into the optical circuit for the CNOT operation as shown in Fig.$\,$4a. The key element in this optical network is a partial polarizing beam splitter (\textit{p}-PBS) which has a transmission of 1(1/3) and a reflectivity of 0(2/3) for the $H$($V$) photons. When the two single photons are superimposed on the \textit{p}-PBS as shown in Fig.$\,$4a, and if one and only one photon leaves through each output channel, the composite state of the two output photons can be written as:
\begin{eqnarray}\label{1}
&& ac\left|H\right\rangle_t\left|H\right\rangle_c  + \sqrt{\frac{1}{3}}\,ad\left|H\right\rangle_t\left|V\right\rangle_c + \sqrt{\frac{1}{3}}\,bc\left|V\right\rangle_t\left|H\right\rangle_c \notag \\
&&+ (\sqrt{\frac{1}{3}}\sqrt{\frac{1}{3}}-\sqrt{\frac{2}{3}}\sqrt{\frac{2}{3}})\,bd\left|V\right\rangle_t\left|V\right\rangle_c
\end{eqnarray}
The first term corresponds to the case in which both input photons are $\left|H\right\rangle$ and fully transmitted. The second and third terms correspond to the cases where one photon is in $\left|H\right\rangle$ and fully transmitted while the other photon is in $\left|V\right\rangle$ and partially (1/3) transmitted. It is most important to note the last term $\left|V\right\rangle_t\left|V\right\rangle_c$, where the resulting minus sign of the probability amplitude ($-$1/3) is due to the quantum interference between two indistinguishable paths, both photons are transmitted ($\sqrt{\frac{1}{3}}\sqrt{\frac{1}{3}}$) or reflected ($-\sqrt{\frac{2}{3}}\sqrt{\frac{2}{3}}$), which requires the indistinguishability of the single photons.

Next, we swap the $H$ and $V$ polarizations in eqn.[\ref{1}] using half-wave plates and pass the two photons through two other \textit{p}-PBSs to compensate the unbalanced coefficient (see Fig.$\,$4a), and we can obtain
$(1/3)(ac\left|H\right\rangle_t\left|H\right\rangle_c+ad\left|H\right\rangle_t\left|V\right\rangle_c+bc\left|V\right\rangle_t\left|H\right\rangle_c-bd\left|V\right\rangle_t\left|V\right\rangle_c).$
This effectively realizes a controlled phase-flip gate with a success probability of 1/9. Finally, after applying two additional Hadamard rotations, it can be transformed into the CNOT gate (see the caption of Fig.$\,$4a and ref. \cite{Kok-RMP,photonCNOT4} for more details).

We experimentally evaluate the performance of the quantum CNOT gate using an efficient method proposed by Hofmann \cite{hofmann}. To show the quantum behaviour of the CNOT gate, it is tested for different combinations of input-output states using complementary bases, that is, in both the computational basis ($\left|0\right\rangle/\left|1\right\rangle$) and their linear superpositions ($\left|\pm\right\rangle=(1/\sqrt{2})(\left|0\right\rangle\pm\left|1\right\rangle)$), which are refereed to as the $ZZ$ and $XX$ basis using the Pauli matrix language respectively. In the $ZZ$ basis, the CNOT is expected to flip the target qubit if the control qubit is in logic 1. Interestingly, in the $XX$ basis, the target and control qubits are reversed: the control qubit will flip if the target qubit is logic 1. The measurement results of the input-output probabilities of the CNOT gate in the $ZZ$ basis and in the $XX$ basis are shown in Fig.$\,$4b and Fig.$\,$4c respectively. The fidelity of the CNOT operation, defined as the probability of obtaining the correct output averaged over all four possible inputs, is in the $ZZ$ basis: $F_{zz}=0.85(6)$, and in the $XX$ basis: $F_{xx}=0.85(7)$. These two complementary fidelities, $F_{zz}$ and $F_{xx}$, are sufficient to give an upper and a lower bound for the full quantum process fidelity $F_{proc}$ of the gate by $(F_{zz}+F_{xx}-1)\leqslant F_{proc} \leqslant min(F_{zz},F_{xx})$. Thus, here we have $0.70(9)\leqslant F_{proc} \leqslant 0.85(7)$. The process fidelity is directly related to the quantum entangling capability of the CNOT gate, that is, the CNOT gate can produce entangled states from unentangled input states \cite{hofmann}. Here, the $F_{proc}$ well surpasses the the threshold of 0.5, which is sufficient to confirm the entangling capability of our CNOT gate.

\subsection{Conclusion and outlook}
In this work, we have demonstrated the on-demand generation of near background-free ($\thicksim$99.7$\%$ purity) and highly indistinguishable RF single photons, from a quantum dot in a planar microcavity driven by resonant $\pi$ pulses. Using two RF photons emitted in 2-ns succession, non-postselective HOM two-photon interference has revealed near-unity visibilities ($\thicksim$97\%), and a quantum CNOT gate with entangling capability has been successfully demonstrated.

Such a pulsed RF single-photon source may open the way to new interesting experiments in quantum optics and quantum information. With the high degree of indistinguishability of the RF photons shown here, they can be used to realize various optical quantum computing algorithms \cite{teleportation,DJ}, interference of multiple photons \cite{Pan-RMP}, and the on-demand generation of photonic cluster state from a single QD \cite{Terry09}. In parallel, the RF spectra of a two-level system under strong pulsed laser excitation which are expected to exhibit novel features beyond the Mollow triplet \cite{pulsedRF} is in itself a subject worth studying.

A natural extension is to realize non-postselective high-visibility quantum interference between two pulsed RF single photons from separate QDs \cite{Glen-PRL,Patel-Nature Photonics}. Based on this, it is possible to entangle remote, independent QD spins \cite{Kimble-Nature2008, Barrett}. We note that although the relatively slow spectral diffusion and pure dephasing does not affect the two-photon interference in our present work due to the 2-ns time separation of the photons, it will limit the degree of indistinguishability for photons from independent QDs. For future experiments, gate-controlled QDs could be used to reduce the spectral diffusion. Alternatively, spectral filtering at the expense of photon rate may be needed.

For quantum information applications, the photon extraction efficiency is a critical issue. So far, we have obtained $\pi$-pulse excited single photons with an overall collection efficiency of 1.3\% reaching the single-photon detector. The photon extraction efficiency can be improved, for example, by embedding the QDs in micropillars or photonic crystal cavities \cite{Nature-Ultrabright2010France}. Large Purcell effects from these microcavities can be helpful to efficiently funnel the spontaneous emission into a guided mode, to further mitigate the dephasings \cite{footnote1}, and increase the pulse repetition rate to tens of GHz. Lastly, it is important to note that in the previous pulsed above-bandgap or \textit{p}-shell excitation experiment, the photon coherence time had to be much larger than the incoherent carrier relaxation time jitter (about tens of ps) in order to obtain a good two-photon interference visibility \cite{footnote1}, which fundamentally put a limit on the radiative lifetime shortening through the Purcell effect. We emphasize that the true resonant, time-jitter-free, pulsed RF technique developed here has no such limitation and can be fully compatible with large Purcell factors to be implemented in the future.

\vspace{0.1cm}
\textit{Acknowledgement}: We thank Y. Yu, Z. Xi, J. Bowles, K. Chen, C. Matthiesen, X.-L. Wang,  L.-J. Wang, N. Vamivakas, and Y. Zhao for helpful discussions. This work was supported by the National Natural Science Foundation of China, the Chinese Academy of Sciences and the National Fundamental Research Program (under Grant No: 2011CB921300, 2013CB933300), and the State of Bavaria. M.A. acknowledges the CAS visiting professorship. C.-Y.L acknowledges the Anhui NSF and Youth Qianren Program.

\vspace{0.15cm}
\textbf{Author contributions}: M.A., C.-Y.L. and J.-W.P. conceived and designed the experiments, C.S., S.H., and M.K. grew and fabricated the sample, Y.-M.H., Y.H., Y.-J.W., D.W., M.A., and C.-Y.L. carried out the optical experiments, Y.-M.H., S.H., C.-Y.L., and J.-W.P. analyzed the data, C.-Y.L. wrote the manuscript with input from all authors, S.H., C.-Y.L. and J.-W.P. guided the project.

\textbf{Additional information:}
The authors declare no competing financial interests.

Correspondence and requests for materials should be addressed to C.-Y.L. (cylu@ustc.edu.cn) or S.H. (sven.hoefling@physik.uni-wuerzburg.de) or J.-W.P. (pan@ustc.edu.cn).

\section*{Supplementary Information}
\renewcommand{\thefigure}{S\arabic{figure}}
 \setcounter{figure}{0}
\renewcommand{\theequation}{S.\arabic{equation}}
 \setcounter{equation}{0}
 \renewcommand{\thesection}{S.\Roman{section}}
\setcounter{section}{0}


\begin{figure*}[htb]
    \centering
        \includegraphics[width=.688\textwidth]{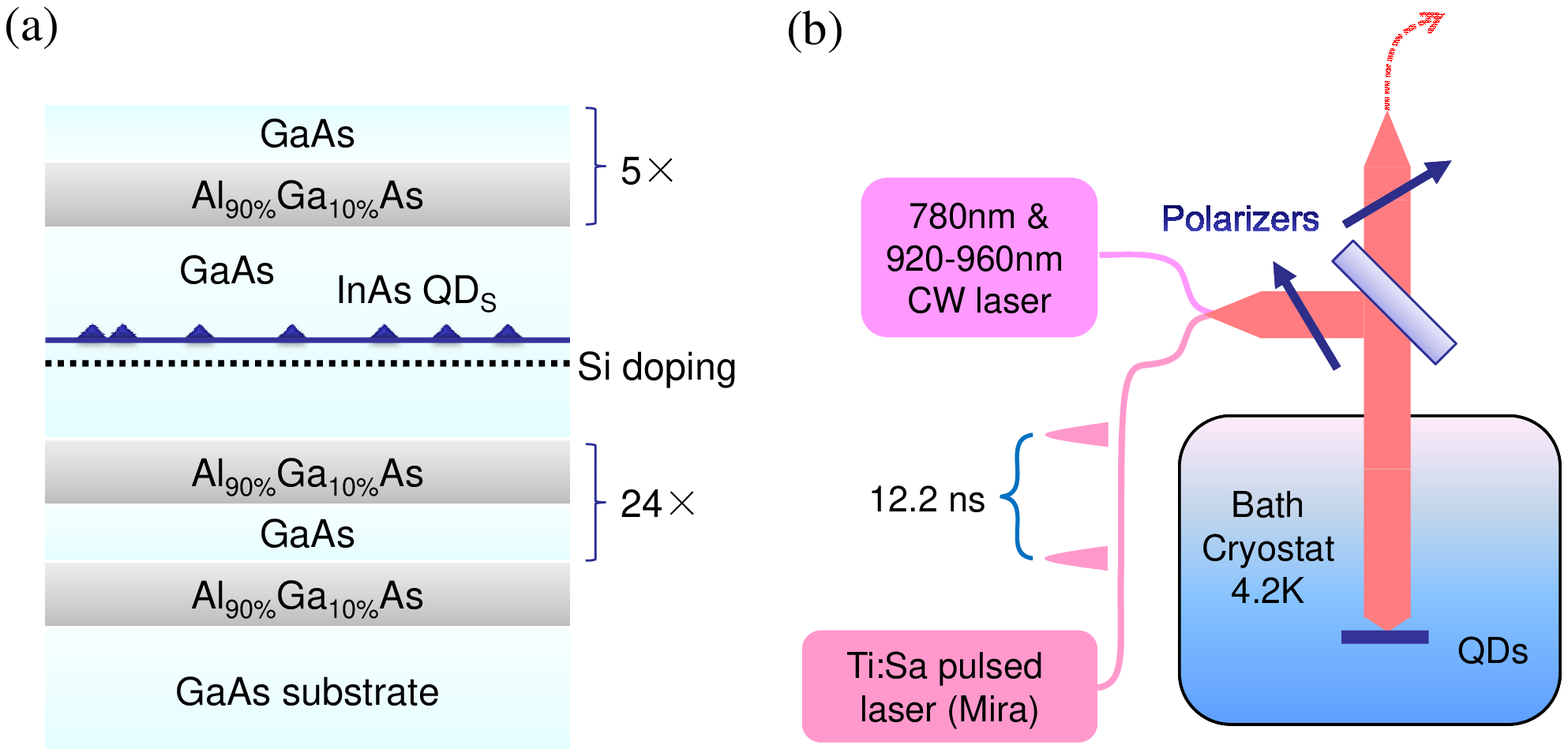}
\caption{The layer structure of the QD sample and the experiment setup. (a). The QD sample is grown by molecular beam epitaxy. The lower and upper distributed-Bragg-reflector mirrors contain 24 and 5 pairs of Al$_{0.9}$Ga$_{0.1}$As/GaAs $\lambda$/4 layers, respectively. The dot density is $\thicksim\,$20/$\mu$m$^2$. (b). The optical excitation of the QD from above band-gap, at \textit{p}-shell, and \textit{s}-shell use CW lasers at wavelength $\thicksim\,$780nm, $\thicksim\,$920.5nm, and $\thicksim\,$941.12nm, respectively. For pulsed excitation, a Ti-sapphire laser (Mira) is used to generate 3-ps optical pulses every 12.2ns. The lasers are power stabilized ($<\,$4\% fluctuation) and focused on the QD sample with an aspheric objective lens (NA=0.68) placed inside a cryogenic-free cryostat (Attodry1000). Single-photon fluorescence was collected through the same lens and directed to single-photon detectors (dark counts $\thicksim\,$50Hz, efficiency $\thicksim\,$22$\%$).}
\label{fig:1}
\end{figure*}

\subsection{Comparison of spectral linewidth}

\begin{figure*}[tb]
    \centering
        \includegraphics[width=.658\textwidth]{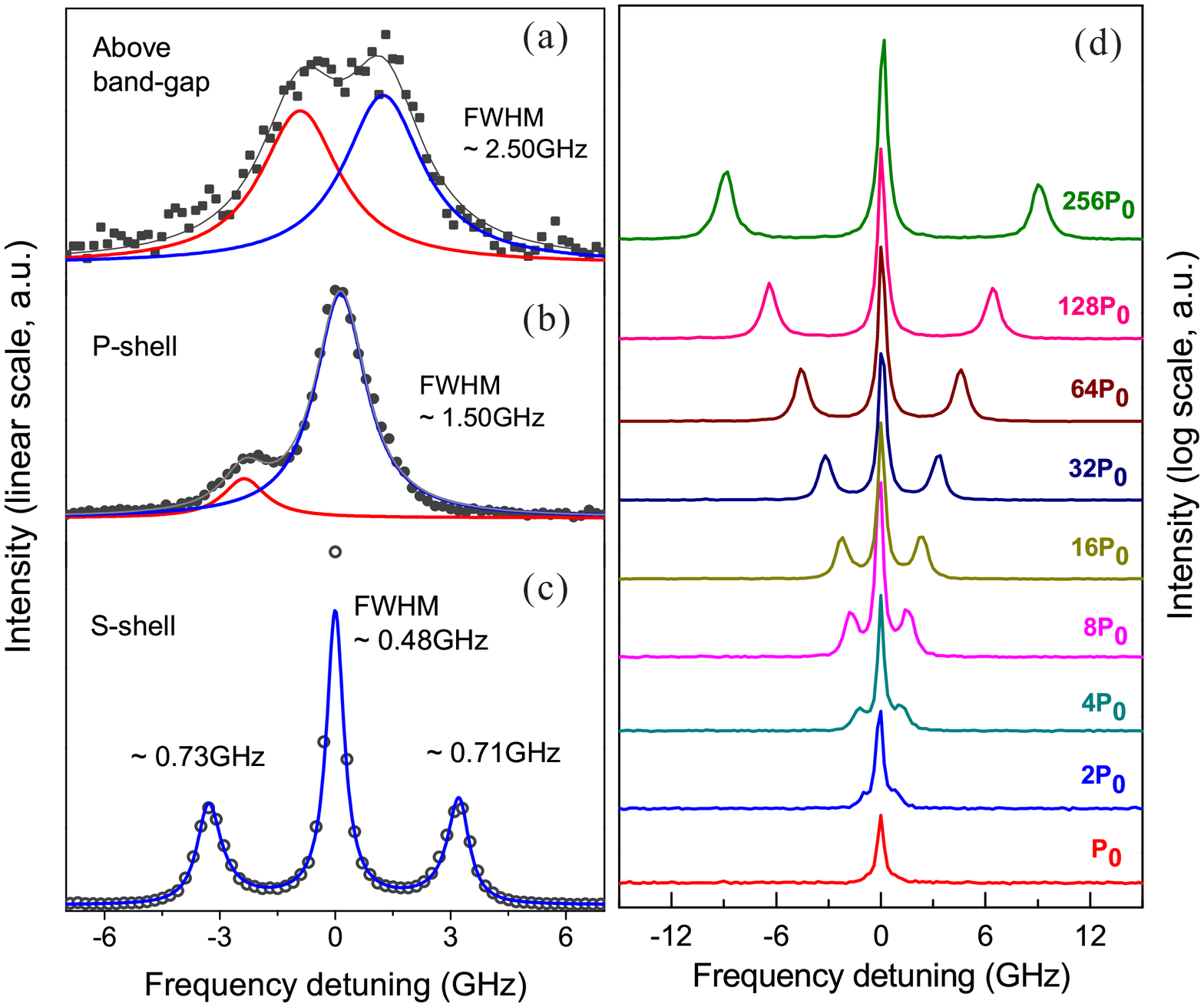}
\caption{Comparison of the emission spectra under different CW excitation methods. To resolve the fine spectral structure, a home-built concave-mirror Fabry-P\'{e}rot cavity with a linewidth of 62MHz, a finesse of 392, and a transmittance of 75\% is used to record the light intensity as a function of cavity transmission frequency. (a) and (b) show the high-resolution spectra of the photon emission from above band-gap and p-shell excitation with CW lasers, respectively. A fine-structure splitting of $\thicksim\,$2.3GHz is visible from the spectra. (c). one example of well-separated Mollow triplet at 32$P_0$ ($\Omega\sim7.8\Gamma_1$) where $\Gamma_1$ is the QD exciton spontaneous emission rate. (d). The RF spectra for a range of excitation laser power ($1.4\Gamma_1<\Omega<22\Gamma_1$). All the data reported are raw data without subtraction of background or smoothing.}
\label{fig:2}
\end{figure*}

In our experiment, we begin with an optical characterization of the QDs and observe a significantly reduced spectral linewidth of the emitted photons from a resonantly driven single QD compared with incoherent excitation methods including via above-bandgap and p-shell using CW lasers. Figure$\,$S2(a-c) present a direct comparison of the spectral linewidth of the emitted photons from a single QD (QD2) neutral exciton for different CW-laser excitation methods. At moderate power regime (around saturation), the CW photoluminescence spectra arising from above band-gap and \textit{p}-shell excitation yields a linewidth of $\sim\,$2.5GHz (see Fig.$\,$S2a) and $\sim\,$1.5GHz (Fig.$\,$S2b), respectively. On the other hand, CW RF photons (see Fig.$\,$S2c) exhibit a significantly narrower linewidth of $\sim\,$0.48GHz even at high power regime well above saturation (32$P_0$) where a Mollow triplet arises \cite{Nick-NatPhys,Flagg-NatPhys,Michler-NatPhotonics}. Figure$\,$S2d shows a series of CW RF spectra at different laser power. The coherence time $T_2$ fitted (using the corrected Eqn.(1) from ref.\cite{Clemens-PRL}) from the CW RF spectra at $P_0$ is closest to being radiative lifetime limited: $T_2$/$2T_{1}$=0.93(6), where $T_1$ is the exciton lifetime which is measured separately to be  of 390(10)ps using time-resolved pulsed RF. This is consistent with the prediction that the pure \textit{s}-shell resonant excitation can eliminate dephasings associated with the incoherent excitation methods \cite{Bennett-OptLett,footnote1}.


\begin{figure*}[bth]
    \centering
        \includegraphics[width=0.5\textwidth]{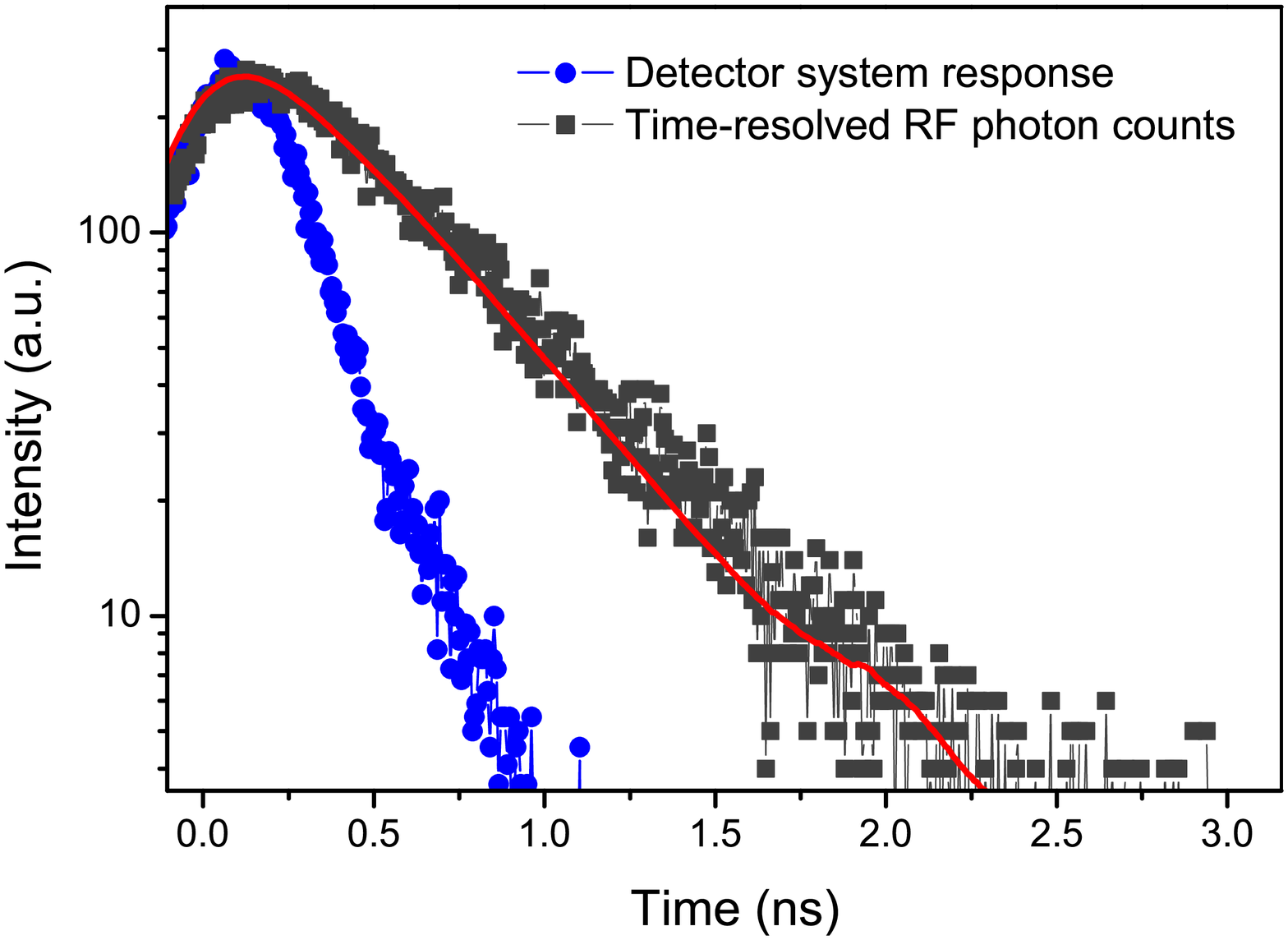}
\caption{Spontaneous emission lifetime of the QD1 single-charged exciton measured by time-resolved pulsed RF. The blue dots are the instrument response function,
the black squares are the measured raw data and the red curve is the best fit to the measured data obtained by convolving the instrument response function with an
exponential function with a decay time of 416(23) ps.}
\label{fig:lifetime}
\end{figure*}

A high-resolution pulsed RF spectrum from QD2 is shown in Fig.$\,$S3. For a range of laser power from 0.2$\pi$ to $\pi$ pulse, we fit the RF spectra with the Voigt profile to extract the inhomogeneous (Gaussian) linewidth ($\omega_G$). As plotted in the inset of Fig.$\,$S3, the $\omega_G$ shows a increase at larger excitation power, which is in qualitative agreement with previous investigations of light-induced spectral diffusion \cite{diffusion1}.

\begin{figure*}[tb]
    \centering
        \includegraphics[width=0.508\textwidth]{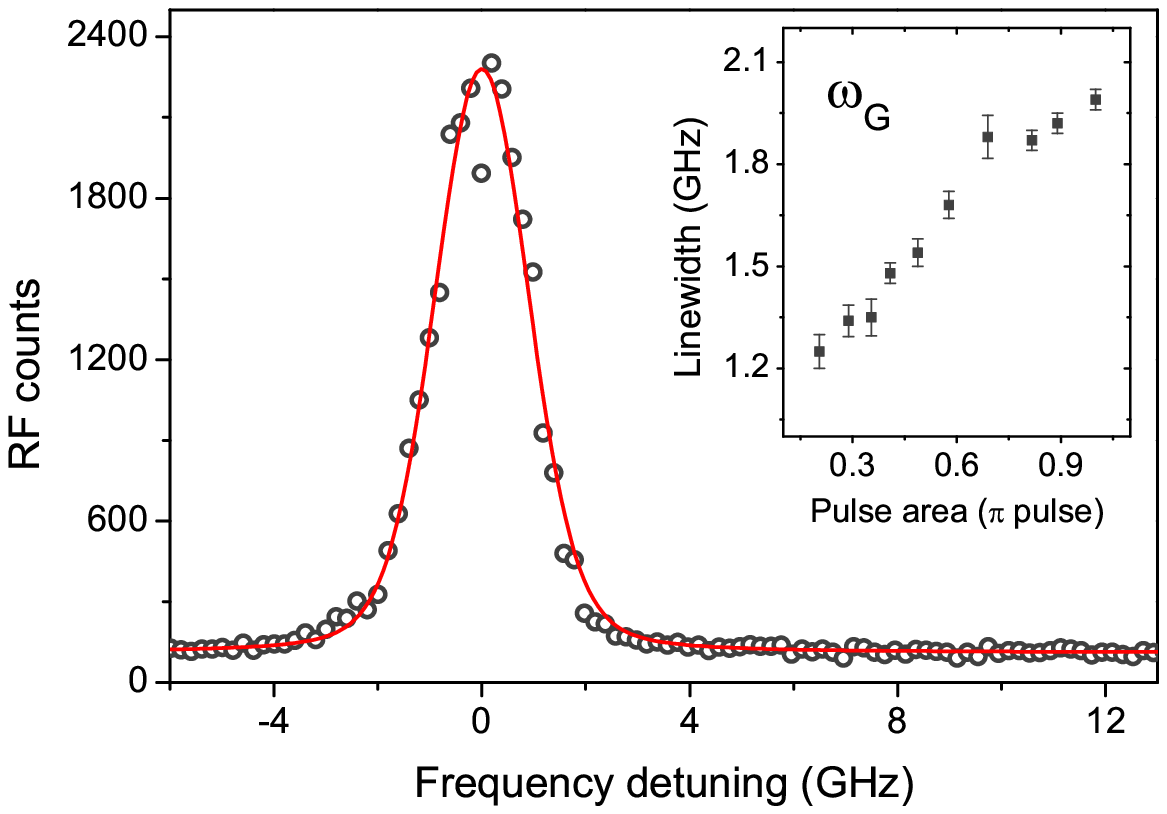}
\caption{A high-resolution RF spectrum when excited by a 0.8$\pi$ pulse. The inset shows the extracted inhomogeneous linewidth by fitting the spectra using a Voigt profile for a range of excitation laser powers.}
\label{fig:3}
\end{figure*}

\begin{figure*}[htb]
    \centering
        \includegraphics[width=0.748\textwidth]{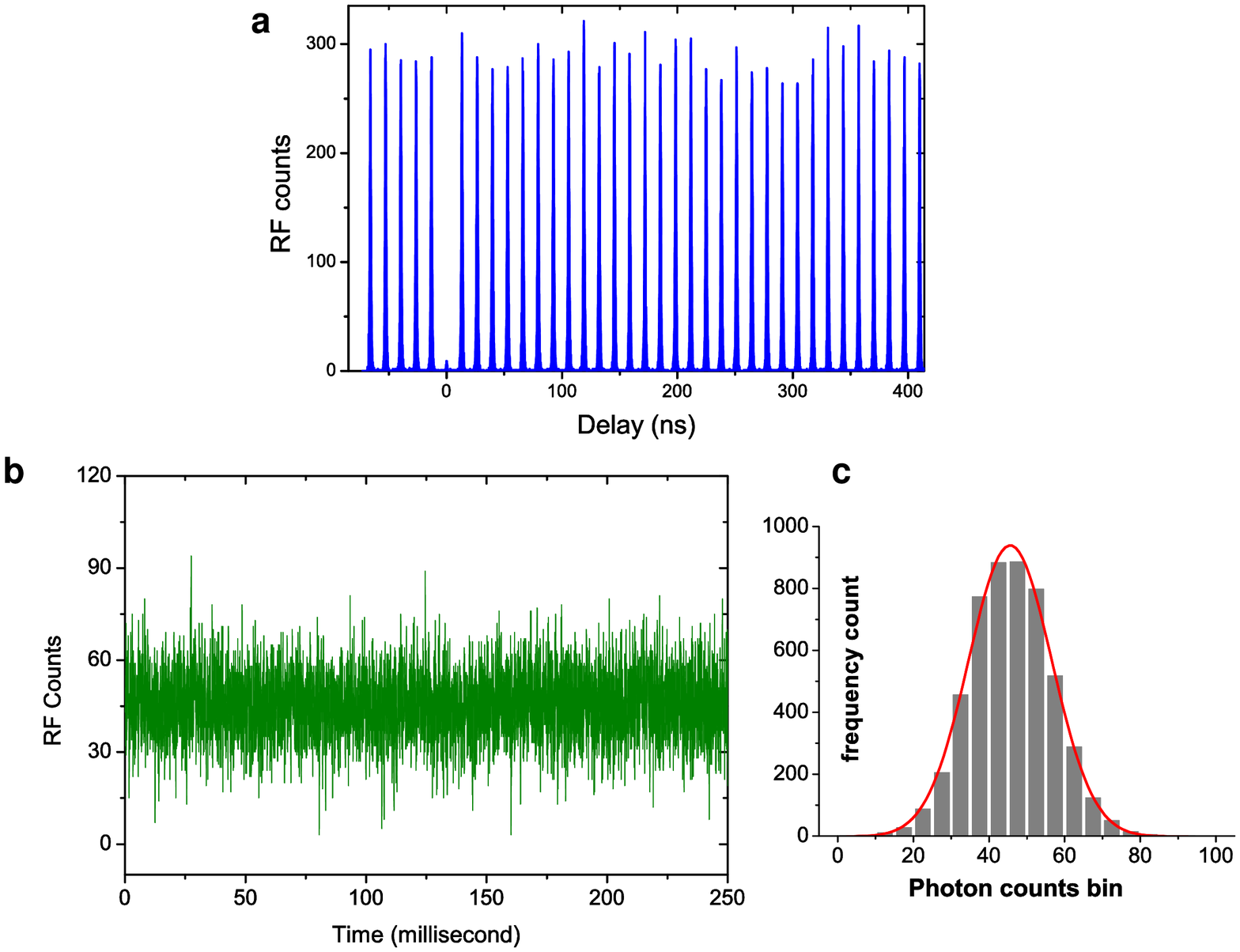}
\caption{Experimental data confirming there is no blinking in our quantum dots. \textbf{a}. Second-order correlation of pulsed RF on a time scale of $\sim$500ns. \textbf{b}. Real time trace of RF with a time bin of 50$\mu$s and the corresponding statistical histogram in \textbf{c}.}
\label{fig:YY}
\end{figure*}

\subsection{HOM interference at different excitation power}

Figure$\,$S4a-b show the data of full histogram obtained on QD2. Clusters of five peaks appear periodically with repetition period of $\thicksim$12.2 ns. The central cluster shows an overall reduced photon counts compared to the side clusters due to the single-photon nature of the source.

The HOM interference are tested with $\pi$, 0.72$\pi$ and 0.41$\pi$ pulse excitation where the RF counts reach $\thicksim\,$$100\%$, $\thicksim\,$$90\%$ and $\thicksim\,$$60\%$ saturation level, and show raw visibilities of 0.903(55), 0.912(56) and 0.934(39), respectively (see Fig.$\,$4(c-h)). Taking into account of the residual two-photon emission probability for this QD, $g^2(0)=0.008(2)$, and the optical imperfections of our interferometric setup (same as in the main text), we obtain corrected degrees of indistinguishability to be 0.956(58), 0.966(59), 0.989(41) for the $\pi$, 0.72$\pi$ and 0.41$\pi$ pulses, respectively.

\begin{figure*}[htb]
    \centering
        \includegraphics[width=0.98\textwidth]{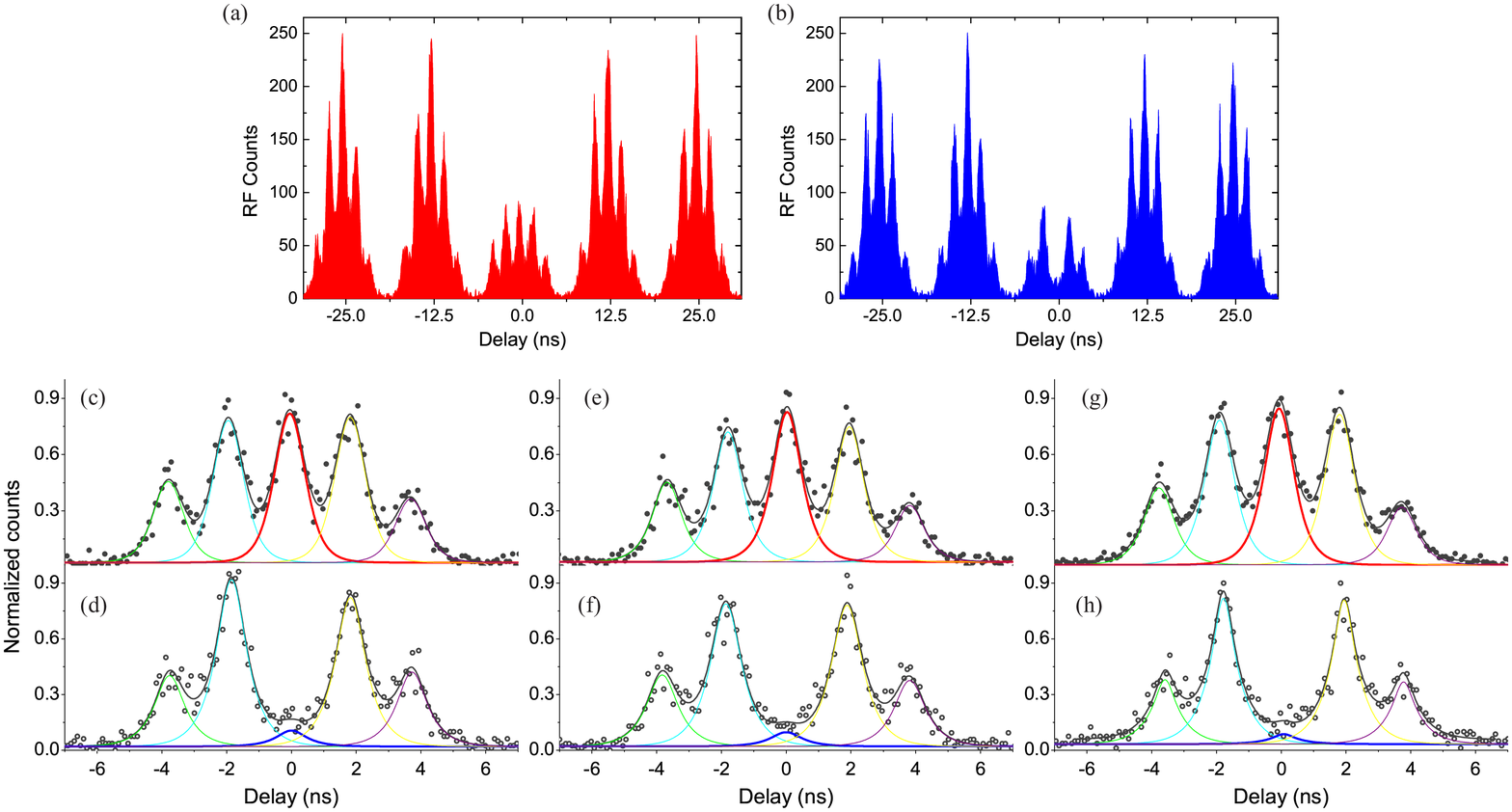}
\caption{The histogram of two-photon detection events with a relative delay time. In (a) and (b), the input two photons are $\pi$-pulse excited and prepared in cross and parallel polarizations, respectively. (c) and (d) are close-ups of the central-cluster feature of (a) and (b). (e-f) and (g-h) are at lower excitation powers at 0.72$\pi$ and 0.41$\pi$ pulse, respectively. The fitting function for each peak is the convolution of a double exponential decay (exciton decay response) with  a Gaussian (single-photon detector time response). Due to the limited time response, the five peaks have finite overlaps. The area of the central peaks (under the red and blue profiles) are extracted and used to calculate the visibility.}
\label{fig:S4}
\end{figure*}


\begin{figure*}[htb]
    \centering
        \includegraphics[width=0.748\textwidth]{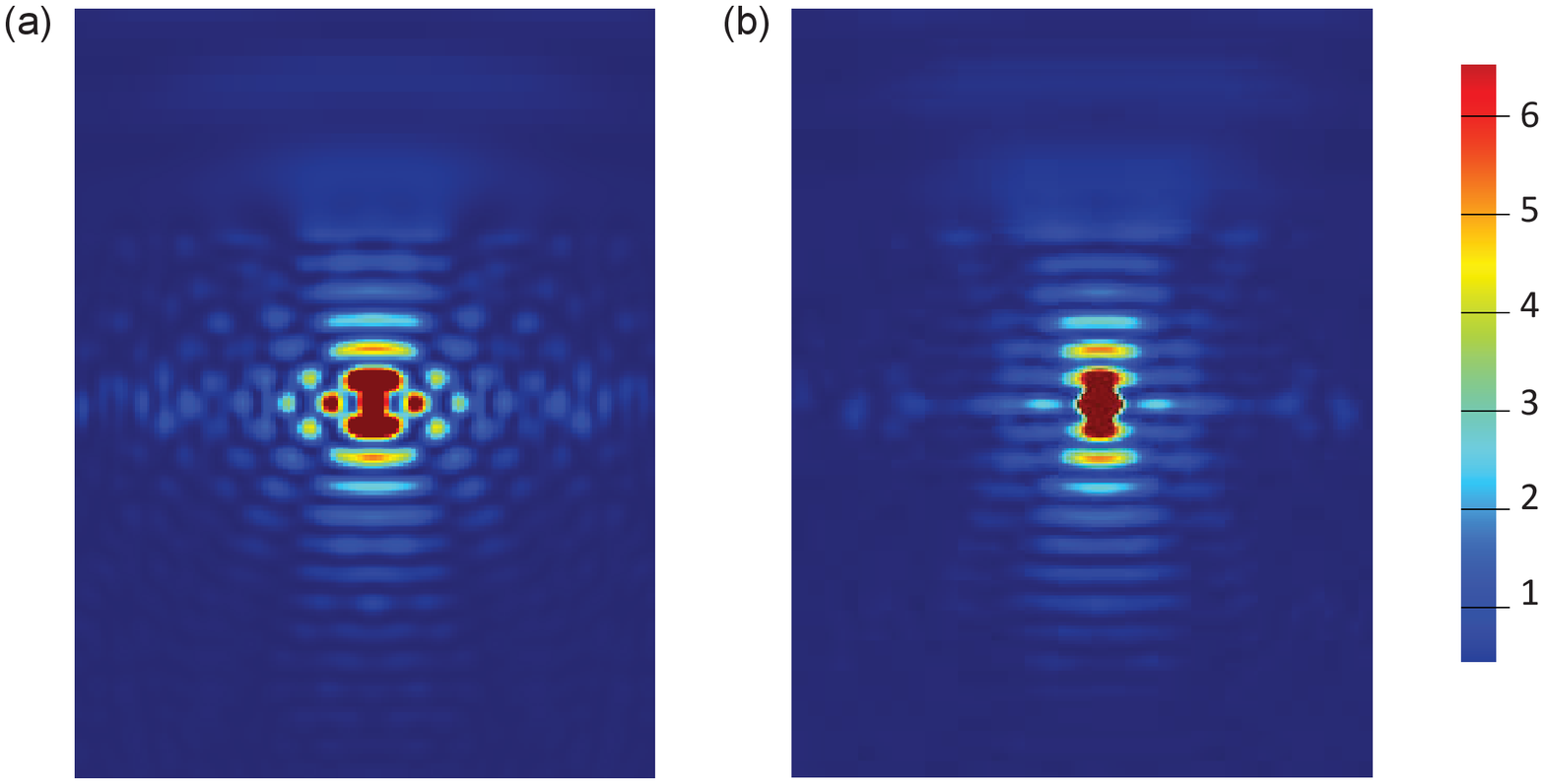}
\caption{A finite-difference time-domain (FDTD) simulation to evaluate the extraction efficiency theoretically based on our QD sample structure as shown Fig.1S. In both in (a) and (b), the y axis is the QD growth direction and the x axis is the x and y plane of the QD respectively. The scale is 4 {$\mu$}m$\times$4{$\mu$}m. The simulation shows that $\sim$14.6$\%$ of the generated photons escaped from the upper GaAs surface, of which $\sim$6.9$\%$ are coupled into the NA=0.68 objective. Experimentally, we excite the QD with $\pi$ pulses at a repetition rate of $\thicksim\,$82MHz and observed $\thicksim\,$230,000 photon counts on a single-photon detector. After correcting for the detection efficiency (22$\%$), the fibre coupling efficiency ($\thicksim\,$45$\%$), polarizer ($\thicksim\,$50$\%$) and beam splitter ($\thicksim\,$95$\%$), we estimate that $\thicksim\,$6$\%$ of the photons emitted by the QD are collected into the first lens, which is in good agreement with the FDTD simulation. The remaining mismatch may have small contributions from the fibre transmission and insertion loss.}
\label{fig:YY}
\end{figure*}

\end{document}